\documentclass[11pt]{article}
\usepackage{paspconf}
\usepackage{wrapfig}
\usepackage{graphicx}

\newcommand{\kms}{{\,\mathrm{km/s}}}

\begin{document}

\title{Dynamical analysis of elliptical galaxy halos}
\author{Andi Kronawitter (1), Ortwin E. Gerhard (1),\\ R. P. Saglia (2), 
	Ralf Bender (2)}
\affil{(1) Astronomisches Institut Universit\"at Basel, \\
       (2) Universit\"atssternwarte M\"unchen}


\section{Introduction}

We report on a project to determine the distribution of dark matter
and the anisotropy of the stellar velocity distribution in elliptical
galaxies, from absorption line profile shapes. These kinematical data
are now available up to 2 -- 3 effective radii.  In the case of the E0
galaxy NGC 6703, the first analysed by our group (Gerhard et al.\
1998), we found that at 2.6 $R_e$ the true circular velocity is $250
\pm 40 \kms$ at $95\%$ confidence, and that correspondingly the
integrated $M/L_B$-ratio in $2.6 R_e$ is at least $\Gamma_B \ge 5.3$,
whereas the central value is $\Gamma_B = 3.3$. The velocity anisotropy 
in this galaxy changes from nearly isotropic in the center to radially
anisotropic at $\sim R_e$, and is not well--determined at $2R_e$,
where $-0.5<\beta<0.5$ for the allowed range of potentials. In this
paper we report some results on the second galaxy investigated, the cD
galaxy in Fornax, NGC 1399.

\section{Data}

In this analysis, photometric data are used to estimate the stellar
mass distribution, and kinematic data are used to model the
distribution of gravitating mass including dark matter. The
photometric data are a composite of HST- (Lauer et al. 1995), CCD-
(Bicknell et al. 1989) and photographic data (Schombert 1986) and go out to
75 $R_e$ for $R_e = 42''$ (Faber et al.\ 1989). Since NGC 1399 is
nearly round (E1) we assume spherical symmetry. In this case the
deprojection of the smoothed surface brightness profile is unique.
Velocity dispersions $\sigma$ and Gauss-Hermite-moments $h_4$ were
measured as described by Bender et al.\ (1994) to $2.3 R_e$, from
spectra obtained using EMMI and the NTT.

\section{Analysis}

The goal is to determine the best spherical potential and phase-space
distribution function (DF) for NGC 1399, modelling both sets of data
in a $\chi^2$ sense.  In practice, we start with a potential
consisting of a luminous and a dark component, determine the best DF
for this potential, and then vary the potential of the dark halo until
the range of suitable potentials is determined. We use an approximate
isothermal sphere model for the dark halo potential; models of this
kind maximize the contribution of the luminous mass ({\em maximum
stellar mass models}).

\section{Results}

\begin{wrapfigure}{r}{8.0cm}
\vspace*{-1.3cm}
\includegraphics[trim=0 20 0 20,width=8.0cm]{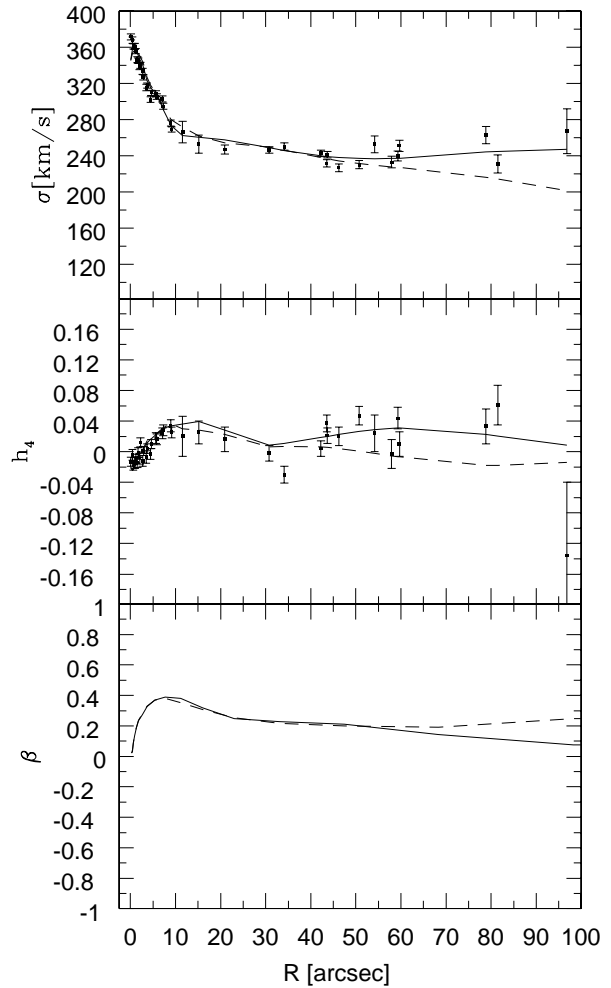}
\caption{Comparison of constant $M/L$ model (dashed) and a model with
a dark halo (full line), to the measured velocity dispersion (top) and
$h_4$--parameter profiles (middle panel).  The bottom panel shows the
intrinisic anisotropy $\beta$ inferred from both models.}
\label{fig1}
\vspace*{-1.3cm}
\end{wrapfigure}
Fig.~\ref{fig1} compares two models to the observed kinematics: one in
which mass follows light, and a second with a suitable dark halo
component.
At large radii, both the velocity dispersion and the $h_4$ profile of
the self-consistent model fall below the data -- this is the signature
of extra mass in the outer parts (Gerhard 1993).  Both profiles are
fit well by the halo model; in the case shown the halo core radius is
$160''$ and the inferred circular velocity at the outermost data point 
is $v_0 = 390 \kms$. The resulting integrated M/L at $2.3 R_e$ is 
$\Gamma_B = 12.4$, whereas the central value is $\Gamma_B = 10.3$.
In either model, the anisotropy parameter $\beta$ is positive, i.e. the
galaxy is radially anisotropic. This confirms the trend emerging from
recently published results (Rix et al. 1997, Gerhard et
al. 1998, Matthias, this conference).


\begin{references}
\reference Bender, R., Saglia, R. P., Gerhard, O. E.  1994, MNRAS, 269, 785
\reference Bicknell, G., Carter, D., Killeen, N., Bruce, T., 1989, 
ApJ, 336, 639
\reference Faber, S. et al. 1989, ApJS, 69, 763
\reference Gerhard, O. E., 1993, MNRAS, 265, 213
\reference Gerhard, O. E., Jeske, G., Saglia, R. P., Bender, R. 1998, 
	MNRAS, 295, 197 
\reference Lauer, T. et al., 1995, ApJ, 110, 2622
\reference Merritt, D. 1993, ApJ, 413, 79
\reference Rix, H.-W., de Zeeuw, P. T., Cretton, N., van der Marel, R.,
	 Carollo, M. C. 1997, ApJ, 488, 702
\reference Schombert, J. M. 1986, ApJS, 60, 603
\end{references}
\end{document}